\documentclass[journal,twoside,web]{IEEEtran}
\usepackage{tmi}
\usepackage{cite}
\usepackage{amsmath,amssymb,amsfonts}
\usepackage{graphicx}
\usepackage{textcomp}
\usepackage{times}
\usepackage{epsfig}
\usepackage{multirow}
\usepackage{tabularx, makecell, booktabs}

\usepackage{enumitem}
\usepackage{hyperref}
 \usepackage{url}
\usepackage{array}
\DeclareUnicodeCharacter{2212}{-}

\def\BibTeX{{\rm B\kern-.05em{\sc i\kern-.025em b}\kern-.08em
    T\kern-.1667em\lower.7ex\hbox{E}\kern-.125emX}}
\begin{document}
\title{A Self Supervised StyleGAN for Image Annotation and Classification with Extremely Limited Labels}

\author{Dana Cohen Hochberg, Hayit Greenspan~\IEEEmembership{Member,~IEEE}, Raja Giryes~\IEEEmembership{Member,~IEEE}
\thanks{D.C. Hochberg is with the School of Electrical Engineering, Tel-Aviv University, Tel-Aviv 6997801, Israel (email: danacohen2@mail.tau.ac.il)}
\thanks{H. Greenspan is with the School of Biomedical Engineering, Tel-Aviv University, Tel-Aviv 6997801, Israel (email: hayit@eng.tau.ac.il)}
\thanks{R. Giryes is with the School of Electrical Engineering, Tel-Aviv University, Tel-Aviv 6997801, Israel (email: raja@tauex.tau.ac.il)}
\thanks{This work was supported by the Ministry of Science and Technology,
Israel. The work of RG is supported by ERC StG under Grant
757497.}
\thanks{2023 IEEE. Personal use of this material is permitted.
  Permission from IEEE must be obtained for all other uses, in any current or future
  media, including reprinting/republishing this material for advertising or promotional
  purposes, creating new collective works, for resale or redistribution to servers or
  lists, or reuse of any copyrighted component of this work in other works.
  DOI: \href{https://doi.org/10.1109/TMI.2022.3187170}{10.1109/TMI.2022.3187170}}
}

\maketitle

\begin{abstract}
The recent success of learning-based algorithms can be greatly attributed to the immense amount of annotated data used for training. Yet, many datasets lack annotations due to the high costs associated with labeling, resulting in degraded performances of deep learning methods. Self-supervised learning is frequently adopted to mitigate the reliance on massive labeled datasets since it exploits unlabeled data to learn relevant feature representations. In this work, we propose SS-StyleGAN, a self-supervised approach for image annotation and classification suitable for extremely small annotated datasets. This novel framework adds self-supervision to the StyleGAN architecture by integrating an encoder that learns the embedding to the StyleGAN latent space, which is well-known for its disentangled properties. The learned latent space enables the smart selection of representatives from the data to be labeled for improved classification performance.
We show that the proposed method attains strong classification results using small labeled datasets of sizes 50 and even 10. We demonstrate the superiority of our approach for the tasks of COVID-19 and liver tumor pathology identification.

\end{abstract}


\section{Introduction}
\label{sec:introduction}
\IEEEPARstart{D}{eep} learning achieved great success in various computer vision tasks, and particularly in supervised learning tasks such as classification. This success is attributed to the large amounts of labeled training data that enable the network to learn meaningful feature representations. Unfortunately, it is often difficult to obtain a satisfactory amount of labeled images since obtaining them is both expensive and time-consuming. Moreover, in the medical field, annotation of medical images additionally requires domain expertise. Therefore, it is often the case in medical image analysis, where there are extremely small labeled datasets with positive pathology cases \cite{frid2018gan,zhou2021review}. This has led to an increase in the search for approaches that perform well with limited annotations.

To overcome this problem current research includes methods such as self-supervised learning \cite{simclr, moco},  few-shot learning \cite{snail_few_shot,meta_few_shot} and active learning \cite{VAAL,coreset}. Active learning addresses the above limitation by querying and annotating the most informative samples from the unlabeled data to achieve the highest classification performance at the lowest labeling cost. Traditionally, active learning is an iterative process where a model is updated in each iteration, and points are selected to be labeled from the unlabeled data in accordance with a set of heuristics. Active learning methods can generally be divided into three main approaches: uncertainty-based\cite{AL_dropout,AL_yoo}, representation-based \cite{coreset}, and their combination \cite{AL_combined_adaptive,AL_combined_cost}. Uncertainty-based methods label samples from the unlabeled data pool that the model is least confident about, hence enhancing the models’ performance when adding them to the labeled pool. Representation-based approaches label the most representative samples of the unlabeled data, thus increasing the diversity of the labeled data pool.

Self-supervised learning has gained considerable popularity in recent years demonstrating promising results on a broad range of computer vision tasks. The advantage of these techniques is that they leverage unlabeled images from the target data domain during a pretraining phase that learns relevant representations of the data, which boosts the performance of various learning tasks \cite{simclr,11_1_gidaris2018unsupervised,14_1_goyal2019scaling,moco,mocov2,49_1_wei2018learning}.

Some works use Generative Adversarial Networks (GANs) in conjunction with self-supervised tasks \cite{4_1_chen2019self,11_1_gidaris2018unsupervised,lt-gan} and establish promising results in unconditional image generation. In an unsupervised manner, GANs learn the feature representations of the data by generating images from input latent codes while training in an adversarial manner. The input latent code acts as the feature representation of the generated image since it contains the necessary information for synthesizing the image. To obtain this representation, an encoder can be added as self-supervision. Adding self-supervision to GANs was shown to improve both the image quality and training stability \cite{27_1_luvcic2019high}. It is generally used as a regularization mechanism for the discriminator which consequently aids the generator in producing higher quality synthesis and better capturing the global structures\cite{gmm-gan-yuri,51_1_zhang2019aet}.

The resolution and quality of synthetic images generated by GANs have significantly improved in recent years. StyleGAN, introduced by Karras et al. \cite{karras2019style,stylegan2} proposes a novel style-based generator that achieves state-of-the-art performance in high-resolution image synthesis. Aside from its unprecedented generation capabilities, StyleGAN learns an intermediate latent space which was shown to contain disentangled properties and enable control over the synthesis process \cite{karras2019style}. These motivated many researchers to embed into the latent space of StyleGAN \cite{abdal2019image2stylegan,hochberg2021style,pidhorskyi2020adversarial,richardson2020encoding} for manipulation of its latent space and various image-to-image translation tasks \cite{abdal2019image2stylegan,richardson2020encoding}.

In this work, we introduce Self-Supervised StyleGAN (SS-StyleGAN), a self-supervised feature representation learning strategy for image annotation and classification that requires minimal labeled data. We present a novel framework that seamlessly combines StyleGAN with an encoder that learns the embedding to its intermediate latent space and leverages its semantically meaningful structure for selecting representatives for labeling and classification. The encoder is incorporated within the StyleGAN architecture thereby adding self-supervision to the framework while learning the feature representations of the data. Consequently, by regularizing the discriminator, the generator is prompted to produce a higher quality synthesis. Once trained, all images are mapped to their latent representations where we employ a smart labeling scheme based on the farthest point sampling algorithm (FPS) \cite{FPS} to select distinct representatives for labeling to achieve high classification performance. 

An important advantage of our approach compared to previous self-supervised methods is that the semantic structure of the latent space induced by StyleGAN allows us to replace the random sampling with the FPS-based smart labeling algorithm that exploits the favorable structure of the latent space. Combining this latent space with our sampling algorithm allows us to obtain superior classification results. Additionally, unlike active learning methods, our approach does not require retraining between each labeling iteration but rather selects all representatives at once. 

Our method is of high medical significance since it allows radiologists to label fewer images while maximizing the classification performance. We validate our method on two medical image classification tasks: COVID-19 and liver tumor classification. We demonstrate the superiority of our method over the current state-of-the-art classification, self-supervised learning, and active-learning approaches.

The main contributions of our work may be summarized as
\begin{itemize}[leftmargin=*]
  \item A new Self-Supervised approach for annotation and classification with very small amounts of labeled data. 
  \item A novel use of the StyleGAN latent space for the task of smart annotation.
  \item Demonstrate the strength of the developed system for the tasks of COVID-19 and liver tumor pathology identification.

\end{itemize}

\begin{figure*}
\begin{center}
\includegraphics[width=\textwidth]{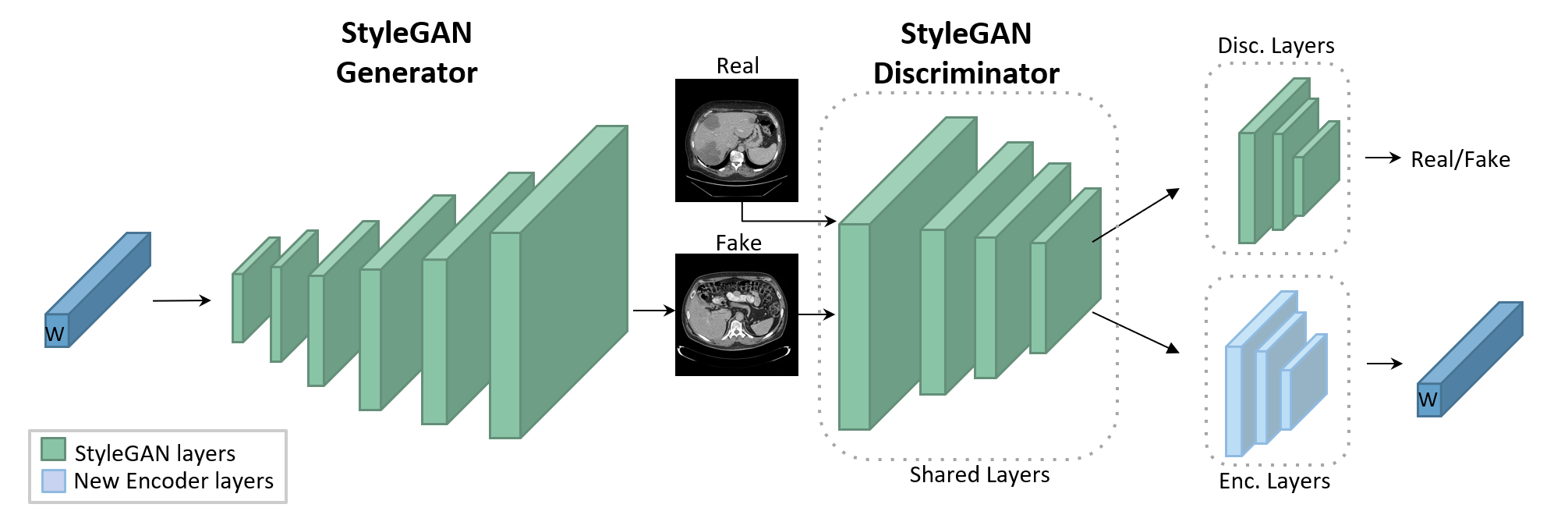}
\end{center}
   \caption{{Proposed SS-StyleGAN framework. Encoder is incorporated within the StyleGAN discriminator  using shared layers which are trained simultaneously with the StyleGAN network. The encoder (with the shared layers) is later used in our annotation and classification framework.}}
\label{framework}
\end{figure*}

\section{Background}
\label{sec:related}
The following section includes an overview of the background and related work. First, we introduce self-supervised learning techniques suitable for training with a limited amount of images. We then describe the StyleGAN latent spaces and strategies for embedding into these spaces. Finally, we present an overview of several active learning approaches, some of which are used as baselines in our experiments.

\subsection{Self-Supervised Learning}
Many existing deep learning approaches are often limited by the lack of annotations, as they lead to poor performance due to over-fitting and biased results. 
Self-supervised learning methods offer a  solution that eliminates the necessity of labeled data by learning the underlying structure of the data while solving some auxiliary pretext task from the input data itself. These representations could then be fine-tuned with a few labels for a supervised downstream task such as image classification, object detection, semantic segmentation, etc.

Recently, contrastive-loss-based learning methods have gained popularity in self-supervised computer vision tasks and achieve impressive performance for learning with small amounts of annotated data \cite{moco,mocov2,simclr}. Instance-discrimination techniques like MoCo \cite{moco,mocov2} and SimCLR \cite{simclr} apply contrastive learning to the entire image instance with the objective of keeping the learned representations invariant under different types of image augmentations. By adding a linear classifier, they achieved classification accuracy that approached the fully supervised learning methods. These approaches have been leveraged in the medical domain to dramatically improve label efficiency for semi-supervised learning \cite{chen2020big,self-supervised-medical}. Some works adapted contrastive learning to the medical domain \cite{sowrirajan2020moco,zhou2020comparing}, while others designed task-specific pretexts \cite{spitzer2018improving,bai2019self}.  For example, Sowrirajan et al. \cite{sowrirajan2020moco} proposed an adaptation of MoCo for improving the classification of chest X-ray models while Azizi et al. \cite{self-supervised-medical} demonstrated the effectiveness of contrastive self-supervised approaches as a pretraining strategy for medical image classification.

\subsection{StyleGAN}
A considerable improvement has been made in image quality and resolution since GANs were first introduced by Goodfellow et al. \cite{goodfellow2014generative}. In recent years, progressive and style-driven approaches have been proposed, setting new standards for image synthesis. StyleGAN \cite{karras2019style,stylegan2} extends the concept of the progressive growing GANs for synthesizing images by continuously increasing the image resolution throughout the training \cite{karras2018progressive}. StyleGAN introduces a novel generator architecture that yields state-of-the-art results in high-resolution image synthesis and provides a new technique for controlling the synthesis process. 

StyleGAN is comprised of several latent spaces. The generator learns a mapping network $M$ which embeds an input latent code $z \in Z$ to a vector in the intermediate latent space $w \in W$, aka "style". $W$ defines the styles that are integrated within the generator architecture. While the distribution of the input vector $z$ is fixed, the distribution of $w$ is learned from the training data itself throughout the training. Therefore, there are no explicit constraints on the structure of $W$ and it learns to capture and disentangle the inherent structure of the training data \cite{karras2019style}. As a result, $w$ latent vectors are more semantically  meaningful than $z$ \cite{yang2021semantic}.

In recent years, multiple studies have explored embedding real-world images into the latent space of GANs. Embedding into these spaces enables the control of the synthesis and therefore opens the door to many image-to-image translation tasks. GAN inversion is used to invert real images back into their latent representations and reconstruct the image via a pretrained generator. Leading approaches for the inversion task include latent vector optimization and using an encoder to map images to the latent space. 
iGAN \cite{iGAN} obtains the embedding of an image by continuous optimization while BiGAN \cite{BiGAN} adds an encoder network to the GAN architecture. This framework is simultaneously trained with the discriminator's objective to classify whether the latent code is real, from a generated image or a real image with an encoded code.  However, DCGAN \cite{dcgan} is used as a Generator in both works, resulting in limited quality. 
Feigin et al. \cite{gmm-gan-yuri} proposed GAEL, a generic architecture that embeds an encoder within the discriminator network with the purpose of creating a latent code from the discriminator’s input images. They demonstrated an improvement in the generated images in both Vanilla GAN and Wasserstein GAN and proved their superiority over BiGAN and other similar methods. %

For higher resolution images, many works use StyleGAN for the embedding due to its high quality synthesis as well as its semantically meaningful latent space which can be utilized for many image-to-image translation tasks \cite{abdal2019image2stylegan,richardson2020encoding,e4e_encoder,face_stylegan_enc,pidhorskyi2020adversarial}. These works typically embed to the $W$ space \cite{stylegan2} or to the extended latent space $W+$ \cite{abdal2019image2stylegan,richardson2020encoding} which is a concatenation of different $w$ vectors, one for each scale of the generator architecture. Note that inverting into $W$ or $W+$ was found to be easier than into an $Z$ and achieved better reconstructions and editing \cite{abdal2019image2stylegan,stylegan2}.

Works embedding into the StyleGAN latent space include embedding either by direct optimization approaches \cite{abdal2019image2stylegan}, encoder-based methods \cite{hochberg2021style,richardson2020encoding,e4e_encoder,pidhorskyi2020adversarial}, or their combination \cite{guan2020collaborative}.
Image2StyleGAN \cite{abdal2019image2stylegan} optimizes a separate style for each scale by projecting the image into the extended latent space $W+$. 
Pidhorskyi et al. proposed ALAE \cite{pidhorskyi2020adversarial}, a StyleGAN-based, progressively growing autoencoder architecture, where the encoder is trained alongside the StyleGAN generator to generate latent codes in the $W$ space. Furthermore, Tov et al. introduced e4e \cite{e4e_encoder}, an encoder designated for the task of image editing by mapping to a latent code comprised of a series of style vectors with a similar distribution of $W$.

Optimization methods typically lead to better reconstruction quality than encoder-based methods, but they require significantly more time. Nonetheless, existing encoder-based methods require the addition of an encoder network which adds computational cost as well as additional training time due to separate training of StyleGAN and the encoder.

While these works focus on the tasks of image reconstruction and manipulation, we leverage the properties of the StyleGAN latent space for an entirely different task of image annotation and classification. Yet, we embed directly into the W latent space which contains the semantic content and maintains the same class as the input image.

Inspired by GAEL \cite{gmm-gan-yuri}, we integrate an encoder within the StyleGAN architecture for the task of self-supervised learning. In this combined framework, the latent space is focused to be suited for inversion while maintaining its disentangled properties and thereby enabling us to encode images into the latent space for the classification task. While the work  in \cite{gmm-gan-yuri} focused on improving image generation quality, our work focuses on self-supervised learning and classification. Our unique combination of the encoder with StyleGAN that both exploits its semantically meaningful learned latent space and allows projecting an image directly into this space provides us with a very strong tool for self-supervision. 
Moreover, compared to existing StyleGAN encoders \cite{pidhorskyi2020adversarial,e4e_encoder}, our encoder is integrated in the discriminator and requires only very little additional computational overhead or training time.

\subsection{Active Learning}
In recent years, several approaches for active learning have been introduced. These techniques can be partitioned into either pool-based or query synthesis methods. Pool-based algorithms select the most informative samples based on multiple sampling strategies. However, in query-synthesis methods, generative models are used to generate the most informative samples rather than querying them from the unlabeled data \cite{AL_synthesis1}. As pool-based approaches are more pertinent to our research, we will focus our review on those studies.

Pool-based methods select new training samples from a given pool of unlabeled data by evaluating the importance of each image based on a given criterion. These methods can be divided into three main categories: uncertainty-based \cite{AL_dropout,AL_yoo}, representation-based \cite{coreset}, and a combination of the two \cite{AL_combined_adaptive,AL_combined_cost}. 
Uncertainty-based methods select samples from the unlabeled data pool to maximize an uncertainty metric that the model is less confident about. Common uncertainty-based sampling heuristics include entropy, least confidence, and margin sampling techniques \cite{AL_sampling}. 
Gal et al. \cite{AL_dropout} use multiple forward passes with Monte Carlo Dropout for estimation of the uncertainty.
Yoo et al. \cite{AL_yoo} propose a Learning-Loss loss method
that attaches a loss prediction module to a task-learner to predict the losses of the unlabeled samples. Hence, the losses are predicted for the entire unlabeled pool and the samples with the top-K losses are labeled.

Representative-based approaches select the highest-diversity data points that will increase the diversity of the labeled pool. To this end, the representation of the data is extracted from the model where the distribution is computed.  
Some works include optimizing the selection of the data by imposing a diversity constraint \cite{AL_diversity} and clustering to select the most representative samples \cite{AL_clustering}.
The Coreset algorithm \cite{coreset} determines representative samples by minimizing the distance between labeled and unlabeled data with intermediate feature information of a trained deep neural network. Nevertheless, Coreset’s optimization algorithm does not scale well as the number of classes, and unlabelled samples increases. Moreover, distance-based representation methods, such as Coreset, do not cope well with high-dimensional data \cite{AL_high}.

Hybrid approaches combine uncertainty and representative-based methods where the samples with the highest uncertainty are selected as the most representative samples in a batch. 
Recently, variational autoencoders (VAE) have been used in conjunction with adversarial training \cite{VAAL} to determine the informativeness of unlabeled samples. Sinha et al. \cite{VAAL} proposed Variational Adversarial Active Learning (VAAL), which trains a VAE alongside an adversarial network to discriminate between the labeled and unlabeled samples.

\section{Method}

Our method consists of several stages: (i) First, an encoder is trained alongside StyleGAN to learn the latent representations of the data; (ii) Next, all images are embedded to the latent space where T-Distributed Stochastic Neighbor Embedding (t-SNE) \cite{tsne} is performed; (iii) The FPS algorithm \cite{FPS} is applied to select the instances for labeling from the unlabeled images that are the most distinct in the latent space; (iv) Finally, The unlabeled images are classified according to the nearest labeled neighbor (NN) in the latent space \cite{nearest_neighbor}.

\begin{figure*}
\begin{center}
\includegraphics[width=\textwidth]{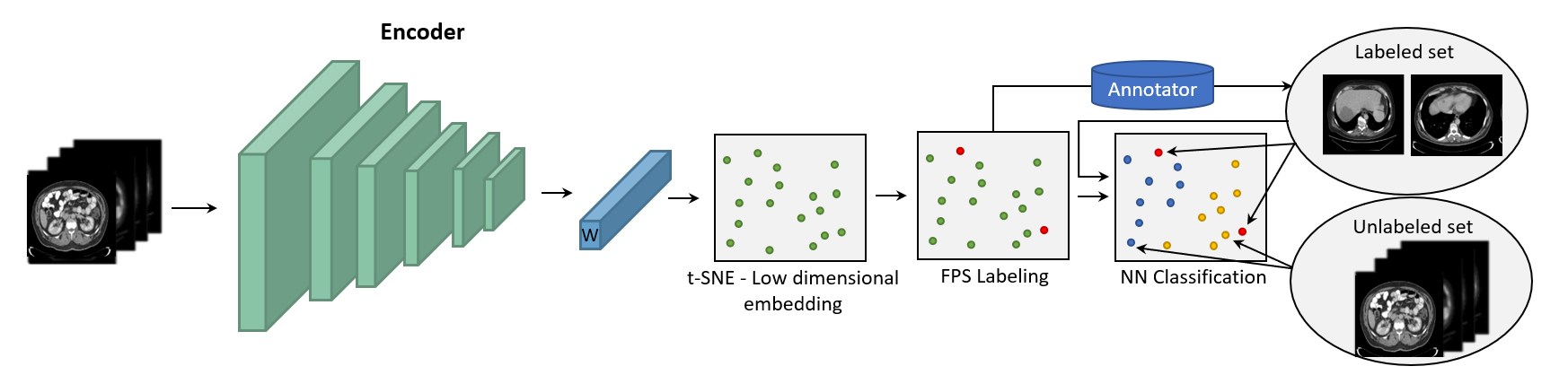}
\end{center}
   \caption{Illustration of the annotation and classification workflow. Images are mapped to their latent representations where t-SNE is performed to compute a low dimensional embedding. Next, FPS is applied to select the representatives to be labeled (red). Finally, the unlabeled images are classified according to their nearest labeled sample.}
\label{class_flow}
\end{figure*}

\subsection{SS-StyleGAN Architecture}

Our proposed architecture is based on the original implementation of StyleGAN2 \cite{stylegan2} with the addition of self-supervision. We incorporate an encoder to the discriminator architecture of StyleGAN2 (noted as StyleGAN for the rest of the paper) that aims to estimate the latent code for the discriminator’s input images. The encoder framework is trained simultaneously with StyleGAN thereby adding self-supervision to the network. The encoder is integrated via shared layers and weights whose output is then forwarded into two small independent similar sub-networks, one for the discriminator and one for the encoder. Fig.~\ref{framework} illustrates our proposed framework. The number of shared layers was determined empirically to achieve optimal results in the training. Refer to Section~\ref{sec:abl_study} for the analysis of the impact of the number of shared layers on the performance of StyleGAN. 

The latent space chosen for the embedding is the intermediate latent space of StyleGAN, $W$, owing to its disentangled properties which we utilize later for both the annotation and classification tasks. StyleGAN is simultaneously trained along with the encoder, therefore, our encoder is able to learn meaningful representations while adding very little training time or computational cost. Our network is trained with a weighted combination of the original StyleGAN loss as well as two additional losses for the encoder. The first loss is derived from the log-likelihood loss in the GAN inversion presented in \cite{gmm-gan-yuri}. The StyleGAN generator, $G$, maps each latent code
$W \in R^M$ from the intermediate latent space to an image $X \in R^{N\times N}$, i.e., $x = G(w)$, and the encoder, $Enc(x)$, embeds an input image $x$ to  the latent space $W$, i.e., $w = Enc(x)$. $\Sigma(x)$ represents an output of the encoder that acts as the variance of the estimation of the encoded images, $Enc(x)$. The second loss includes the mean squared error (MSE) between the original image, $x_{real}$ and reconstructed image $G(Enc(x_{real}))$.  

The overall encoder loss is 
\begin{equation}
\begin{aligned}
\mathcal{L}_{Enc} &= -\lambda E_{x,w\sim P(x,w)}[\log(P(w|x)] \\ 
&+\beta \mathcal{L}_{MSE}(x_{real},G(Enc(x_{real}))),
\label{enc}
\end{aligned}
\end{equation}
where 
\begin{equation}
\begin{aligned}
\log(P(w|x)) &= −0.5\log((2\pi)^{M}|\Sigma(x)|) \\
 & \hspace{-0.2in} -0.5(w - Enc (x))^T\Sigma^{-1}(x)(w − Enc (x)).
\label{gael}
\end{aligned}
\end{equation}

\subsection{Representative-based Nearest Neighbor Classification}
One of the main advantages of the style-based generative framework is its well-behaved, disentangled feature space which can be exploited for downstream tasks. For the task of classification, this is especially profitable as only a few annotations, selected by a smart sampling algorithm, are required to generalize over the entire space.

Once SS-StyleGAN is trained, all images are embedded to the latent space where t-SNE \cite{tsne} and FPS \cite{FPS,FPS_image} are employed. t-SNE enables the representation of high-dimensional data in a low-dimensional space while maintaining both the local and global data structures and emphasizing the similarity between the data points. This attribute along with its non-linearity have enabled t-SNE to surpass other dimensionality reduction techniques, such as PCA (principal components analysis) \cite{PCA}, in various applications including classification \cite{tsne_face,tsne_tumor}. 
We apply t-SNE on the encoder's output, the intermediate 512-dimensional latent vector, to project it to a 2D space.

FPS is then employed for iteratively sampling a set of k representatives for labeling. FPS starts by selecting a random point and iteratively selects the point with the largest geodesic distance to the ones previously selected until $k$ (the number of samples to be annotated) are chosen. The points selected by FPS are labeled and the others are classified based on their nearest labeled sample with a nearest neighbor classifier \cite{nearest_neighbor} applied on the 2D t-SNE space. An illustration of the annotation and classification workflow is provided in Fig. \ref{class_flow}.

\subsection{Datasets}
\begin{figure}
\centering
\includegraphics[width=\linewidth]{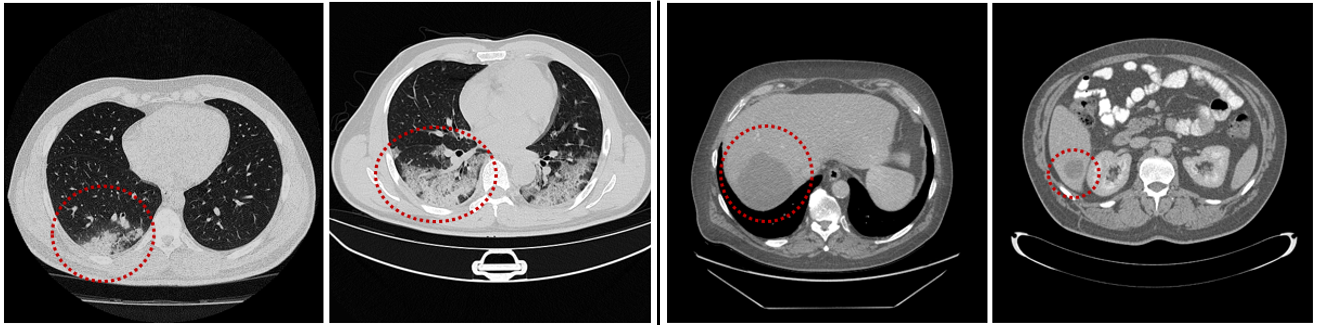}
    \caption{Example images of the COVID-19 (left) and LiTs (right) datasets. Lung opacities (COVID-19) and tumors (LiTs) are circled in red.}
\label{GT_images}    
\end{figure}

To demonstrate our method, we examined two medical imaging datasets. Examples of images from each dataset are presented in Fig. \ref{GT_images}.
\subsubsection{COVID-19 Dataset}
Two small annotated COVID-19 publicly available datasets were combined: The first, MedSeg COVID-19 CT dataset \cite{covid_data_cheng}, contains 9 axial volumetric CT scans with both COVID-19 positive and negative slices. The second dataset \cite{covid_joseph} consists of 20 CT scans that contain patients who are positive or negative but with other types of pneumonia. Axial slices smaller than $0.9$ and larger than $0.22$ were used from both datasets leading to 3,036 slices in total which are divided into COVID-19 positive and negative slices.  {Preprocessing included windowing the Hounsfield unit (HU) values in the range [-1024, 325], resizing the images to a resolution of $512\times512$ and replicating each slice to create the 3 channels of an RGB image.}

\subsubsection{LiTs Dataset}
Computed tomography (CT) scans from the publicly available training dataset of the Liver Tumor Segmentation Challenge (LiTS) challenge \cite{LiTS}. The images are of resolution of $512\times 512$ and contain slices coming from the entire abdomen region which we separated into two groups: slices with tumors in the liver and without, {as we use this dataset for a binary classification task. To maintain only the relevant organs the HU values were windowed to the range [-300, 300]. Additionally, as mentioned, the images were converted to RGB images.}

\subsection{Implementation Details}
We build upon the official TensorFlow implementation of StyleGAN \cite{stylegan2}. We modify the architecture to include an encoder by sharing 12 resolution layers with the discriminator that are followed by 4 additional (non-shared) layers for each. All the training details and parameters are identical to configuration F of StyleGAN. Our models were trained on images of resolution $512\times512$ but could easily be adapted to other resolutions as well. All networks were trained on a single NVIDIA GeForce GTX 1080 Ti GPU.

\section{Experiments and Results}
To assess the effectiveness of our approach on the classification performance we compared it to a state-of-the-art classification network, a self-supervised method, and two active learning methods. We additionally conduct an ablation study to estimate the impact of the self-supervision on StyleGAN, investigate the optimal configuration of the encoder and annotation algorithm, and highlight the importance of the latent space used for the embedding on the classification task.

\subsection{Experimental setup}

We divide our method into two phases: self-supervised training and classification. For the self-supervised phase, SS-StyleGAN was trained in an unsupervised manner on all images from each dataset. For the LiTs dataset, we fine-tuned a StyleGAN pretrained on the FFHQ dataset \cite{karras2019style} and for the COVID-19 training, we fine-tuned StyleGAN pretrained on the LiTs dataset, since the LiTs dataset is much larger in size and has similar anatomical structure. The same training parameters were used for both models. The framework was trained with $\lambda=10$ and $\beta = 1$. Similar to \cite{gmm-gan-yuri}, $\Sigma(x)$ was modeled as the identity matrix, $\Sigma(x) = I$, for increased simplicity. {For the dimension reduction technique in the classification phase, t-SNE was trained for 1000 iterations with a learning rate of 200 and a perplexity of 10 for the COVID-19 dataset and 50 for the LiTs dataset.}

\begin{figure}
\begin{center}
       \includegraphics[width=\linewidth,height=4.5cm] {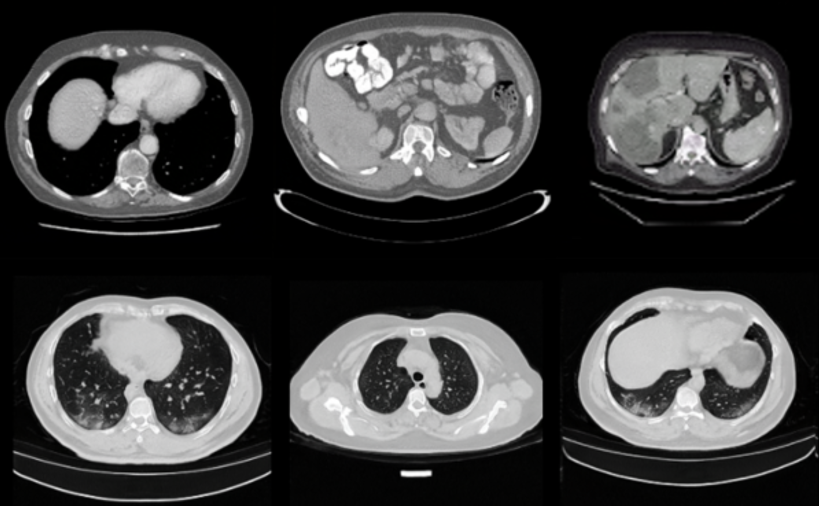} 
\end{center}
    \caption{Example slices of unsupervised image synthesis by SS-StyleGAN for the LiTs (top row) and COVID-19 (bottom row) datasets.} \label{generated}
\end{figure}

In the comparison of the classification task, we consider as baselines the following state-of-the-art frameworks: the EfficientNet classification network \cite{efficientnet}, the MoCo v2 \cite{mocov2} self-supervised learning approach, and two known active learning methods, VAAL \cite{VAAL} and the Learning Loss algorithm \cite{AL_yoo}.

Transfer learning {(TL)} was applied for both EfficientNet and MoCo v2 from ImageNet \cite{deng2009imagenet} as it was shown to improve classification results of medical images for both supervised and self-supervised methods \cite{self-supervised-medical}.  {Furthermore, for the COVID-19 dataset, due to its small size, we trained MoCo v2 on the LiTs dataset prior to fine-tuning on COVID-19, referred to as MoCo v2 (TL) as it has been demonstrated that pretraining MoCo on a medical dataset leads to improved representations and results \cite{moco_pretraining}}. Next, the encoder layers were frozen and a linear classifier was trained as described in \cite{mocov2}.

For a fair comparison, we also employ our representative labeling approach on MoCo v2, referred to as MoCo v2 (FPS). The labeling is applied in the same manner as in our approach, namely using the t-SNE algorithm in the learned latent space of MoCo, followed by FPS and NN classification. Furthermore, we present the results of our method, SS-StyleGAN, with random sampling (RS) instead of FPS.

We compare our model to two active learning approaches; a GAN-based method, VAAL\cite{VAAL}, and Learning Loss \cite{AL_yoo}. Another active learning method tested was the Coreset approach \cite{coreset}. This method was not able to achieve adequate results even when training with 50 images in our experiments and therefore we do not display its results.
The training of the active learning models is initialized with a single labeled image from the training set while the rest of the training images comprise the unlabeled data set. In each iteration, ten images are selected from the unlabeled data for annotation. These images are then added to the labeled training set and the training is repeated.

We trained the active learning models three times and all other models fives times, each with a different train and test set (randomly selected each time), due to the extremely long training time required for adequate training of the active learning methods. 

{Additionally, to asses the generalization ability of our model on other data that has not been used for training SS-StyleGAN, we tested our annotation and classification algorithms on another COVID-19 dataset from China Consortium of Chest CT Image Investigation (CC-CCII) \cite{covid_dataset_unseen}. This dataset is publicly available and contains COVID-19 positive and negative CT scans. For this dataset, the performance of our annotation and classification strategy was evaluated on 722 COVID positive and 294 negative CT slices. 
}

{Note that the encoder is tuned in our model not for accurate reconstruction (as is the case with the StyleGAN inversion models) but rather for improving the generation quality, which is reflected in the higher FID score that is achieved (Table \ref{FID}), and creating a useful representation that can be used for embedding and classification afterwards. Therefore, we do not measure its performance in terms of reconstruction quality in the experiments as it does not represent the functionality of our designed solution.}

\subsection{Ablation Study}
\label{sec:abl_study}
To evaluate the significance of each aspect of our suggested approach, we conducted several experiments to evaluate the effect of self-supervision on the quality of the generated images, find the optimal encoder framework and annotation method, and examine the latent space for the embedding.

The comparison of the different models and configurations was performed by evaluating the quality of the generated images in each. To that end, the Fréchet Inception Distance (FID) metric \cite{FID} was computed. FID is a measure of the difference between two distributions in the feature space of an InceptionV3 classifier \cite{inception} and is often used to assess the quality of the synthesized images by GANs. Table \ref{FID} reports the FID metric of our model in various configurations and of the original StyleGAN model when trained with the LiTs dataset. Results show that our model outperforms StyleGAN, reaching an FID score of 11.6 compared to 16.3. In other words, by incorporating self-supervision, the generator is empowered to a higher quality synthesis. Examples of generated images by our model for both datasets are displayed in Fig. \ref{generated}.

Furthermore, our model offers significantly better performance when trained with 12 shared layers (4 non-shared) in comparison to the other configurations, and was therefore the selected configuration. This performance can be observed in Table \ref{FID} by the decrease in FID scores as the number of shared layers declines and reaches a minimum at 12 shared layers.

\begin{table}[!hbt]
    \caption{FID comparison (lower is better) of different configurations of SS-StyleGAN and StyleGAN for the LiTs dataset. The configurations include a different number of shared layers between the encoder and discriminator.} \label{FID}
    \centering
    \begin{tabular}{c|c|c}
        \toprule 
        \thead{\bf{\normalsize Model}} & \thead{\bf{\normalsize Shared Layers}} & \thead{\bf{\normalsize FID}} \\ \hline
        SS-StyleGAN & 15 &  28.0\\
        SS-StyleGAN & 14 & 23.8\\
        SS-StyleGAN & 13 & 17.3\\
        \textbf{SS-StyleGAN} & \textbf{12} & \textbf{11.6}\\
        SS-StyleGAN & 11 & 16.4\\
        StyleGAN2 & - & 16.3\\
        \bottomrule
\end{tabular}
\end{table}

\begin{table*}[t]
    \caption{The impact of embedding into the $Z$ and $W$ latent space on the classification performance (higher is better). The results are displayed for training with 50 labeled images. }\label{results_ablation}
    \centering
    \makebox[\textwidth][c]{
    \begin{tabular}{c|c|c|c|c|c|c}
    \toprule
    
        \thead{\bf{\normalsize Dataset}} & \thead{\bf{\normalsize Model}} & \thead{\bf{\normalsize Accuracy}} &  \thead{\bf{\normalsize Sensitivity}} & \thead{\bf{\normalsize Specificity}} & \thead{\bf{\normalsize Precision}} & \thead{\bf{\normalsize AUC}} \\ \hline
        LiTs & SS-StyleGAN($Z$ loss, $Z$ space)  & 0.58 $\pm$ 0.02 & 0.57 $\pm$ 0.03 & 0.59 $\pm$ 0.02 & 0.40 $\pm$ 0.03 & 0.58 $\pm$ 0.02\\
        LiTs & SS-StyleGAN($W$ loss, $Z$ space) &  0.56 $\pm$ 0.03 & 0.55 $\pm$ 0.03 & 0.60 $\pm$ 0.07 & 0.38 $\pm$ 0.04 & 0.57 $\pm$ 0.04 \\
        LiTs & SS-StyleGAN & \bf{0.89 $\pm$ 0.02} & \bf{0.89 $\pm$ 0.03} & \bf{0.89 $\pm$ 0.06} & \bf{0.79 $\pm$ 0.04} & \bf{0.89 $\pm$ 0.02}\\ \hline
        
        COVID-19 & SS-StyleGAN($W$ loss, $Z$ space) &  0.72 $\pm$ 0.06 & 0.76 $\pm$ 0.09 & 0.69 $\pm$ 0.05 & 0.75 $\pm$ 0.09 & 0.72 $\pm$ 0.06 \\
        COVID-19 & SS-StyleGAN & \bf{0.92 $\pm$ 0.03} & \bf{0.97 $\pm$ 0.03} & \bf{0.89 $\pm$ 0.05} & \bf{0.97 $\pm$ 0.02} & \bf{0.93 $\pm$ 0.02}\\
        
    \bottomrule
    \end{tabular}}
\end{table*}

We explore the optimal latent space used for the classification task by comparing the performance of our model when embedding into each of the latent spaces of StyleGAN, $Z$, and $W$. Table \ref{results_ablation} presents a comparison of the classification performance when mapping to each space for both datasets. As a reminder, our setup, SS-StyleGAN, includes both embedding into $W$ space and training with a loss on $W$ (Equation \ref{gael}). The results show significantly degraded performance when embedding into $Z$ in all experiments. This does not come as a surprise as $W$ is known for its disentangled properties and meaningfulness.  
Moreover, to address the question of whether the results are derived from the latent space used in the loss throughout the training, we further employ our method with the loss on $Z$ for the LiTs dataset, noted as SS-StyleGAN ($Z$ loss, $Z$ space). As shown in Table \ref{results_ablation}, this configuration does not produce reasonable results as well.

Furthermore, we examined our method with several variations {presented in Table \ref{results_ablation_DR_methods}}; classification directly on $W$ and linear classification instead of NN on $W$ before and after t-SNE. The results in these experiments were inferior to our current setup. 
This confirms that the t-SNE data dimensionality reduction preserves only the essential information leading to improved classification performance.     
Moreover, we evaluated our method with PCA instead of t-SNE {using the aforementioned variations and also in a setup, where we add a Multi-layer Perceptron (MLP) classifier. Note, FPS was applied in all cases to select the images to annotate. All experiments led to degraded performance as can be seen in Table \ref{results_ablation_DR_methods}.} 
Since PCA is a linear technique it is incapable of capturing non-linear dependencies. Moreover, unlike t-SNE, PCA does not preserve the local structures of the data \cite{tsne_tumor} which is essential considering our task.

\begin{table*}[t]
    \caption{{Ablation study for different types of dimension reduction and classification methods for the COVID-19 and LiTs datasets when training with 10, 20 and 50 labeled images. Results displayed are average AUC results over 5 experiments (higher is better). Our chosen configuration is displayed in bold.}  }\label{results_ablation_DR_methods}
    \centering
    \makebox[\textwidth][c]{
    \begin{tabular}{c|c|c|c|c|c|c|c}
    \toprule
    
        \thead{\bf{\normalsize Images}} & \thead{\bf{\normalsize Dimension Reduction Method}} & \thead{\bf{\normalsize Classification Method}} & \thead{\bf{\normalsize AUC (COVID-19)}} &  \thead{\bf{\normalsize AUC (LiTs)}} \\ \hline
        
        10 & - & NN  & 0.57 $\pm$ 0.04 & 0.59 $\pm$ 0.03\\
        
        10 & PCA & NN &   0.61 $\pm$ 0.10 & 0.63 $\pm$ 0.12\\
        
        10 & PCA & MLP &  0.57 $\pm$ 0.08 & 0.70 $\pm$ 0.02\\
        
        10 & PCA & Linear  & 0.63 $\pm$ 0.13 & 0.70 $\pm$ 0.07\\
        
        10 & t-SNE & Linear & 0.77 $\pm$ 0.02 & 0.65 $\pm$ 0.03\\
        
        10 & \bf{t-SNE} & \bf{NN} & \bf{0.82 $\pm$ 0.03} & \bf{0.72 $\pm$ 0.02}\\
        \hline
        
        20 & - & NN  &  0.59 $\pm$ 0.07 & 0.61 $\pm$ 0.03\\
        
        20 & PCA & NN &   0.74 $\pm$ 0.05  & 0.66 $\pm$ 0.03 \\
        
        20 & PCA & MLP &  0.76 $\pm$ 0.06 & 0.75 $\pm$ 0.04 \\
        
        20 & PCA & Linear &  0.79 $\pm$ 0.06 & 0.74 $\pm$ 0.06\\
        
        20 & t-SNE & Linear & 0.79 $\pm$ 0.01 & 0.65 $\pm$ 0.05\\
        
        20 & \bf{t-SNE} & \bf{NN} & \bf{0.87 $\pm$ 0.04} & \bf{0.78 $\pm$ 0.06}\\
        \hline
        
        50 & - & NN  &  0.63 $\pm$ 0.04 & 0.66 $\pm$ 0.04\\
        50 & PCA & NN &   0.82 $\pm$ 0.04  & 0.68 $\pm$ 0.04\\
        50 & PCA & MLP &  0.89 $\pm$ 0.03 & 0.84 $\pm$ 0.04\\
        
        50 & PCA & Linear &  0.89 $\pm$ 0.04 & 0.82 $\pm$ 0.02\\
        50 & t-SNE & Linear &  0.78 $\pm$ 0.03 & 0.69 $\pm$ 0.01\\
        50 & \bf{t-SNE} & \bf{NN} & \bf{0.93 $\pm$ 0.02} & \bf{0.89 $\pm$ 0.02}\\
        
    \bottomrule
    \end{tabular}}
\end{table*}

\subsection{Classification Results}

\begin{figure}[t]
\begin{center}
\includegraphics[width=\linewidth]{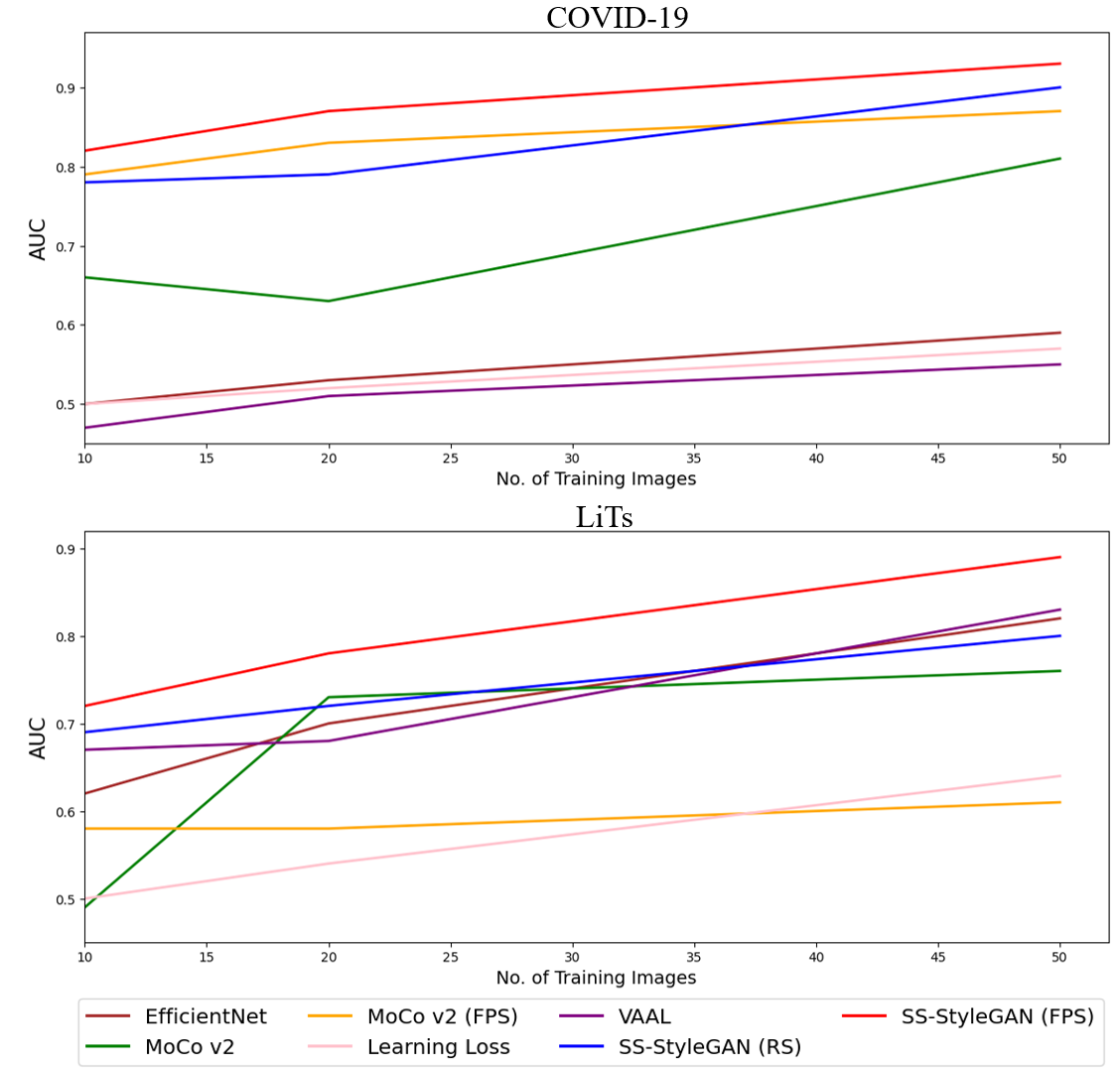}
\end{center}
\caption{Comparison of AUC scores (higher is better) for all models for the COVID-19 (left) and LiTs (right) datasets. The results show that our model outperforms all of the other methods in all experiments.}
\label{covid_lits_plots}
\end{figure}

\begin{figure*}
\begin{center}
       \includegraphics[width=\textwidth,height=3.8cm] {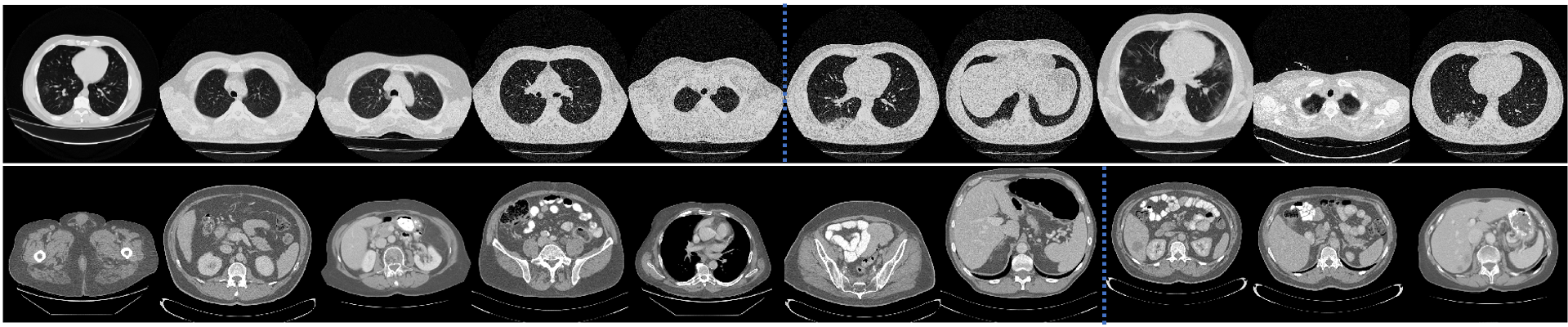} 
\end{center}
    \caption{Images selected by our annotation algorithm for the COVID-19 (top) and LiTs (bottom) training datasets. Images right of the dotted line are pathology positive, left are negative. Notice, samples with large variability from different slices of the lungs/abdomen are selected.  
    } \label{FPS_images}
\end{figure*}

To evaluate our classification performance, we conducted a series of experiments. In each experiment, 15\% of the CTs were selected randomly as the test set and the rest were used for training. From the training set 10, 20, or 50 slices were selected for labeling.  
Table \ref{results_50_COVID} and \ref{results_50_LITS} display the results of the mean and standard deviation of all models for both COVID-19 and LiTs datasets respectively. 

The results confirm that our model is capable of generalizing even from an extremely limited set of training images. With only 10 and 20 images we achieve an AUC of (0.82 $\pm$ 0.03, 0.87 $\pm$ 0.04) and (0.72 $\pm$ 0.02, 0.78 $\pm$ 0.06) for the COVID-19 and LiTs datasets respectively. Our model outperforms all other models for both datasets and was capable of attaining meaningful results even when most others completely failed. For example, for the COVID-19 dataset and 20 images training, EffitientNet fails with an AUC of 0.53 and we outperform the next best model, MoCo v2 (FPS), with a 4\% gain in AUC and the traditional MoCo v2 ({with and without TL from LiTs}), with a gain of {more than 24\%}. Similarly, for both datasets with 50 training images, we outperform all models in AUC by over 6\%. Fig. \ref{covid_lits_plots} demonstrates that our model achieved significantly higher AUC scores than any other model on both datasets. 

Fig. \ref{FPS_images} displays the images selected by our annotation algorithm for the 10 labeled images experiment. The images show that our approach selects samples of different lungs or abdominal heights. Moreover, for the COVID-19 dataset, an equal number of positive and negative slices were chosen. The large variability in the selected examples provides reasoning to the performance achieved with our algorithm.

As previously mentioned, transfer learning was used for most competing models. We can therefore conclude that transfer learning alone is not enough in the scenarios presented. As opposed to all other models, our model is capable of handling small amounts of training data, whereas most other models require more images to ensure adequate training. This task remains very challenging, even for MoCo v2, which is also a self-supervised method, and all active learning models. 

The results indicate that the FPS algorithm improves the performance of our model, but even without it, it remains superior to most others. Therefore, it can be concluded that our model learns meaningful feature representations of the data and that even a small number of images are sufficient for its generalization. All these are unique and set apart our model from other approaches.

\begin{table*}[t]
    \caption{{Average classification results on the test set for the COVID-19 dataset when training with 10, 20 and 50 images from all experiments (higher is better).}}\label{results_50_COVID}
    \centering
    \makebox[\textwidth][c]{
    \begin{tabular}{c|l|c|c|c|c|c}
    \toprule
    
        \thead{\bf{\normalsize Images}} & \thead{\bf{\normalsize Model}} & \thead{\bf{\normalsize Accuracy}} &  \thead{\bf{\normalsize Sensitivity}} & \thead{\bf{\normalsize Specificity}} & \thead{\bf{\normalsize Precision}} & \thead{\bf{\normalsize AUC}} \\ \hline
        10 & EfficientNet &  0.56 $\pm$ 0.04 & 0.51 $\pm$ 0.41 & 0.54 $\pm$ 0.42 & 0.45 $\pm$ 0.25 & 0.50 $\pm$ 0.10 \\
        10  & MoCo v2 & 0.61 $\pm$ 0.04 & 0.44 $\pm$ 0.14 & 0.76 $\pm$ 0.15 & 0.62 $\pm$ 0.03 & 0.60 $\pm$ 0.03 \\
        10 & MoCo v2 (TL) & 0.65 $\pm$ 0.04 & 0.71 $\pm$ 0.21 & 0.61 $\pm$ 0.21 & 0.74 $\pm$ 0.06 & 0.66 $\pm$ 0.03\\
        10 & MoCo v2 (FPS)   & 0.79 $\pm$ 0.04 & 0.75 $\pm$ 0.10 & 0.73 $\pm$ 0.09 & 0.78 $\pm$ 0.08 & 0.79 $\pm$ 0.04\\
        10  & Learning Loss & 0.45 $\pm$ 0.22 & 0.75 $\pm$ 0.42 & 0.26 $\pm$ 0.42 & 0.60 $\pm$ 0.37 & 0.50 $\pm$ 0.03\\
        10  & VAAL & 0.37 $\pm$ 0.12 & 0.74 $\pm$ 0.43 & 0.22 $\pm$ 0.44 & 0.23 $\pm$ 0.33 & 0.47 $\pm$ 0.06\\
        10  & SS-StyleGAN (RS) & 0.79 $\pm$ 0.06 & 0.71 $\pm$ 0.17 & \bf{0.85 $\pm$ 0.11} & 0.81 $\pm$ 0.10 & 0.78 $\pm$ 0.05\\
        10 & SS-StyleGAN (FPS)  & \bf{0.82 $\pm$ 0.03} & \bf{0.86 $\pm$ 0.06} & 0.79  $\pm$ 0.10 & \bf{0.86 $\pm$ 0.04} & \bf{0.82 $\pm$ 0.03}\\\hline        
        20 & EfficientNet &  0.56 $\pm$ 0.07 & 0.60 $\pm$ 0.32 & 0.47 $\pm$ 0.25 & 0.62 $\pm$ 0.22 & 0.53 $\pm$ 0.16 \\
        20 &  MoCo v2 & 0.60 $\pm$ 0.11 & 0.40 $\pm$ 0.15 & 0.84 $\pm$ 0.10 & 0.57 $\pm$ 0.21 & 0.61 $\pm$ 0.07 \\
        20 & MoCo v2 (TL)& 0.67 $\pm$ 0.17 & 0.85 $\pm$ 0.31 & 0.42 $\pm$ 0.43 & 0.48 $\pm$ 0.50 & 0.63 $\pm$ 0.21\\
        20 & MoCo v2 (FPS) & 0.83 $\pm$ 0.02 & 0.77 $\pm$ 0.04 & \bf{0.90 $\pm$ 0.02} & 0.79 $\pm$ 0.06 & 0.83 $\pm$ 0.02\\
        20  & Learning Loss & 0.59 $\pm$ 0.17 & 0.20 $\pm$ 0.44 & 0.83 $\pm$ 0.38 & 0.72 $\pm$ 0.18 & 0.52 $\pm$ 0.04\\
        20  & VAAL & 0.30 $\pm$ 0.03 & \bf{0.99 $\pm$ 0.01} & 0.02 $\pm$ 0.04 & 0.02 $\pm$ 0.03 & 0.51 $\pm$ 0.02\\
        20  & SS-StyleGAN (RS) & 0.79 $\pm$ 0.07 & 0.77 $\pm$ 0.15 & 0.80 $\pm$ 0.10 & 0.83 $\pm$ 0.08 & 0.79 $\pm$ 0.07\\
        20 & SS-StyleGAN (FPS)  & \bf{0.87 $\pm$ 0.04} & 0.94 $\pm$ 0.04 & 0.81  $\pm$ 0.07 & \bf{0.93 $\pm$ 0.04} & \bf{0.87 $\pm$ 0.04}\\\hline
     
        50 & EfficientNet &  0.59 $\pm$ 0.11 & 0.44 $\pm$ 0.37 & 0.70 $\pm$ 0.30 & 0.64 $\pm$ 0.18 & 0.59 $\pm$ 0.20 \\
        50 & MoCo v2 & 0.84 $\pm$ 0.07 & 0.88 $\pm$ 0.12 & 0.79 $\pm$ 0.09 & 0.92 $\pm$ 0.04 & 0.83 $\pm$ 0.08 \\
        50 & MoCo v2 (TL)      & 0.78 $\pm$ 0.21 & 0.91 $\pm$ 0.01 & 0.73 $\pm$ 0.23 & 0.90 $\pm$ 0.04 & 0.81 $\pm$ 0.13\\
        50 & MoCo v2 (FPS) & 0.87 $\pm$ 0.03 & 0.85 $\pm$ 0.06 & 0.89 $\pm$ 0.03 & 0.89 $\pm$ 0.04 & 0.87 $\pm$ 0.03\\
        50  & Learning Loss & 0.68 $\pm$ 0.04 & 0.25 $\pm$ 0.29 & 0.89 $\pm$ 0.12 & 0.72 $\pm$ 0.13 & 0.57 $\pm$ 0.10\\
        50  & VAAL & 0.52 $\pm$ 0.22 & 0.61 $\pm$ 0.53 & 0.49 $\pm$ 0.50 & 0.49 $\pm$ 0.46 & 0.55 $\pm$ 0.10\\
        50  & SS-StyleGAN (RS) & 0.90 $\pm$ 0.03 & 0.87 $\pm$ 0.08 & \bf{0.93 $\pm$ 0.06} & 0.90 $\pm$ 0.07 & 0.90 $\pm$ 0.04\\
        50 & SS-StyleGAN (FPS)  & \bf{0.92 $\pm$ 0.03} & \bf{0.97 $\pm$ 0.03} & 0.89 $\pm$ 0.05 & \bf{0.97 $\pm$ 0.02} & \bf{0.93 $\pm$ 0.02}\\

    \bottomrule
    \end{tabular}}
\end{table*}

\begin{table*}[t]
    \caption{Average classification results on the test set for the LiTs dataset when training with 10, 20 and 50 images from all experiments (higher is better).}\label{results_50_LITS}
    \centering
    \makebox[\textwidth][c]{
    \begin{tabular}{c|l|c|c|c|c|c}
    \toprule
        \thead{\bf{\normalsize Images}} & \thead{\bf{\normalsize Model}} & \thead{\bf{\normalsize Accuracy}} &  \thead{\bf{\normalsize Sensitivity}} & \thead{\bf{\normalsize Specificity}} & \thead{\bf{\normalsize Precision}} & \thead{\bf{\normalsize AUC}} \\ \hline
        10   & EfficientNet & 0.66 $\pm$ 0.03 & 0.76 $\pm$ 0.15 & 0.41 $\pm$ 0.28 & 0.56 $\pm$ 0.24 & 0.62 $\pm$ 0.07\\
        10  & MoCo v2 & 0.59 $\pm$ 0.17 & 0.75 $\pm$ 0.38 & 0.43 $\pm$ 0.42 & 0.25 $\pm$ 0.28 & 0.49 $\pm$ 0.03\\
        10  & MoCo v2 (FPS) & 0.64 $\pm$ 0.08 & 0.74 $\pm$ 0.20 & 0.43 $\pm$ 0.20 & 0.48 $\pm$ 0.10 & 0.58 $\pm$ 0.10\\
        10  & Learning Loss & 0.42 $\pm$ 0.25 & 0.33 $\pm$ 0.57 & 0.67 $\pm$ 0.57 & 0.19 $\pm$ 0.16 & 0.50 $\pm$ 0.01\\
        10  & VAAL & 0.69 $\pm$ 0.18 & 0.72 $\pm$ 0.35 & 0.62 $\pm$ 0.46 & 0.42 $\pm$ 0.33 & 0.67 $\pm$ 0.16\\
        10   & SS-StyleGAN (RS)  & 0.72 $\pm$ 0.06 & \bf{0.77 $\pm$ 0.11} & 0.61 $\pm$ 0.17 & \bf{0.57 $\pm$ 0.07} & 0.69 $\pm$ 0.07\\ 
        10   & SS-StyleGAN (FPS)  & \bf{0.72 $\pm$ 0.03} & 0.71 $\pm$ 0.12 & \bf{0.72 $\pm$ 0.05} & 0.56 $\pm$ 0.05 & \bf{0.72 $\pm$ 0.02}\\\hline
        
        20   & EfficientNet & 0.67 $\pm$ 0.14 & 0.69 $\pm$ 0.21 & 0.67 $\pm$ 0.15 & 0.54 $\pm$ 0.19 & 0.70 $\pm$ 0.15\\
        20  & MoCo v2  & 0.80 $\pm$ 0.06 & \bf{0.96 $\pm$ 0.03} & 0.50 $\pm$ 0.24 & \bf{0.91 $\pm$ 0.07} & 0.73 $\pm$ 0.10\\
        20  & MoCo v2 (FPS) & 0.65 $\pm$ 0.08 & 0.79 $\pm$ 0.18 & 0.38 $\pm$ 0.15 & 0.50 $\pm$ 0.12 & 0.58 $\pm$ 0.02\\
        20  & Learning Loss & 0.51 $\pm$ 0.22 & 0.45 $\pm$ 0.50 & 0.62 $\pm$ 0.54 & 0.45 $\pm$ 0.50 & 0.54 $\pm$ 0.07\\
        20  & VAAL & 0.77 $\pm$ 0.07 & 0.88 $\pm$ 0.09 & 0.48 $\pm$ 0.34 & 0.47 $\pm$ 0.34 & 0.68 $\pm$ 0.15\\
        20   & SS-StyleGAN (RS)  & 0.74 $\pm$ 0.07 & 0.77 $\pm$ 0.07 & 0.67 $\pm$ 0.17 & 0.59 $\pm$ 0.06 & 0.72 $\pm$ 0.02\\
        20   & SS-StyleGAN (FPS)  & \bf{0.80 $\pm$ 0.01} & 0.79 $\pm$ 0.04 & \bf{0.76 $\pm$ 0.17} & 0.62 $\pm$ 0.02 & \bf{0.78 $\pm$ 0.06} \\\hline

        50   & EfficientNet &  0.75 $\pm$ 0.05 & 0.79 $\pm$ 0.11 & 0.66 $\pm$ 0.13 & 0.61 $\pm$ 0.10 & 0.82 $\pm$ 0.04 \\
        50  & MoCo v2 & 0.82 $\pm$ 0.06 & 0.97 $\pm$ 0.07 & 0.55 $\pm$ 0.32 & \bf{0.94 $\pm$ 0.12} & 0.76 $\pm$ 0.16\\ 
        50  & MoCo v2 (FPS) & 0.67 $\pm$ 0.01 & 0.78 $\pm$ 0.04 & 0.43 $\pm$ 0.05 & 0.48 $\pm$ 0.03 & 0.61 $\pm$ 0.01\\
        50  & Learning Loss & 0.78 $\pm$ 0.05 & \bf{0.97 $\pm$ 0.01} & 0.31 $\pm$ 0.18 & 0.76 $\pm$ 0.10 & 0.64 $\pm$ 0.08\\
        50  & VAAL & 0.84 $\pm$ 0.03 & 0.87 $\pm$ 0.13 & 0.80 $\pm$ 0.13 & 0.73 $\pm$ 0.08 & 0.83 $\pm$ 0.04\\
        50   & SS-StyleGAN (RS)  & 0.83 $\pm$ 0.03 & 0.88 $\pm$ 0.06 & 0.72 $\pm$ 0.11 & 0.76 $\pm$ 0.08 & 0.80 $\pm$ 0.04 \\
        50   & SS-StyleGAN (FPS)  & \bf{0.89 $\pm$ 0.02} & 0.89 $\pm$ 0.03 & \bf{0.89 $\pm$ 0.06} & 0.79 $\pm$ 0.04 & \bf{0.89 $\pm$ 0.02} \\

    \bottomrule
    \end{tabular}}
\end{table*}

\subsection{The Learned Latent Space}

\begin{figure}[t]
\begin{center}
\includegraphics[width=9.3cm]{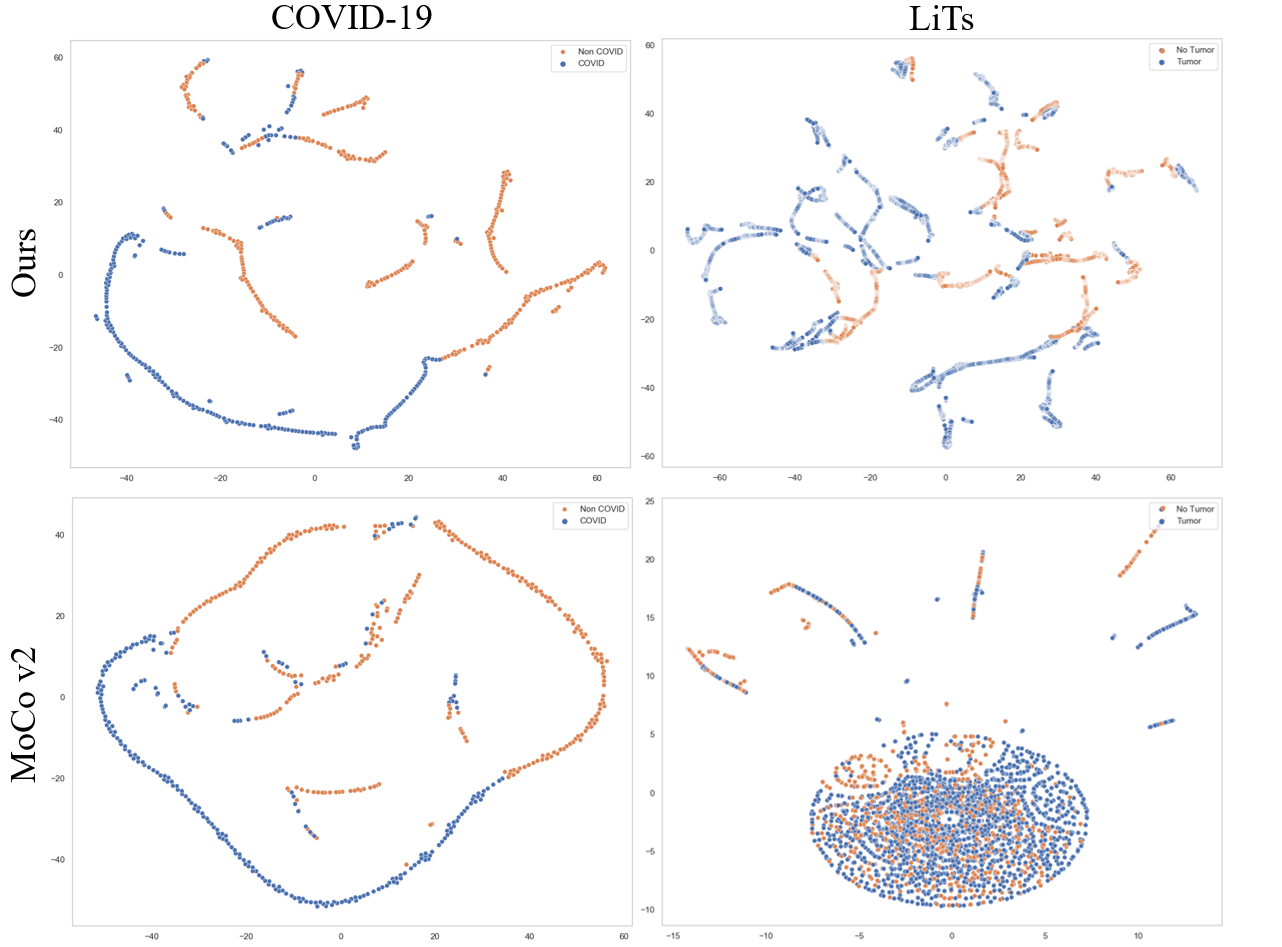}
\end{center}
\caption{Latent space visualization via t-SNE projection for our method (top) and MoCo v2 (down) of 10 CT volumes from the COVID-19 (left) and LiTs (right) datasets. 
By embedding directly into the StyleGAN’s intermediate latent space, we are able to obtain higher separation between the classes than the latent space induced by MoCo v2, where the classes cannot always be distinguished (as can be seen for the LiTs dataset). Thus, the latent space induced by our method is substantially more informative and therefore superior for classification.
}
\label{tsne_covid}
\end{figure}

To better understand the reason behind our method's success, we visualize the t-SNE features of our learned latent space and of MoCo v2 in Fig. \ref{tsne_covid}. The plots indicate the presence of distinct classes for our method in both datasets. Similarly, the classes in the MoCo v2 latent space for the COVID-19 dataset can also be distinguished, which contributes to the improved performance obtained by MoCo v2 (FPS) over the conventional method (Table \ref{results_50_COVID}). However, for the LiTs dataset, MoCo v2 was entirely unable to differentiate between the classes. This confirms that the latent space imposed by our method is capable of capturing the underlying representation of the data and reflects how the classification can be performed with such a limited number of images.

\begin{figure*}
\begin{center}
       \includegraphics[width=16cm,height=4.0cm] {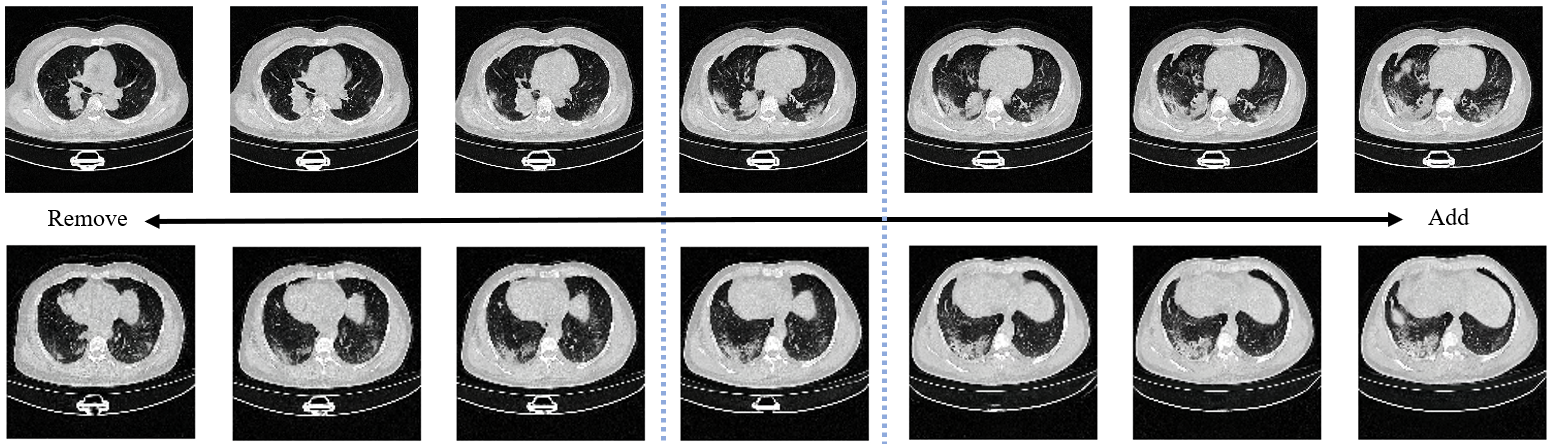} 
\end{center}
    \caption{Latent space manipulation examples (rows) . Adding and removing the COVID-19 opacity from original image (dotted square).} \label{add_remove_covid}
\end{figure*}

To further exhibit the good properties of the learned space, we demonstrate the disentanglement of the class attribute in the latent space by manipulating this attribute. The semantically meaningful latent space of StyleGAN allows us to determine which direction vectors represent individual factors of variation, specifically class direction vectors. By finding the boundary line that separates between the classes in the latent space, and moving towards each side we can affect the extent to which the class will appear in the generated image. Towards this end, we present an example with the COVID-19 dataset. Fig. \ref{add_remove_covid} displays the manipulation of images by adding and removing lung opacities associated with this disease.

\begin{figure}[t]
\begin{center}
\includegraphics[width=8.8cm]{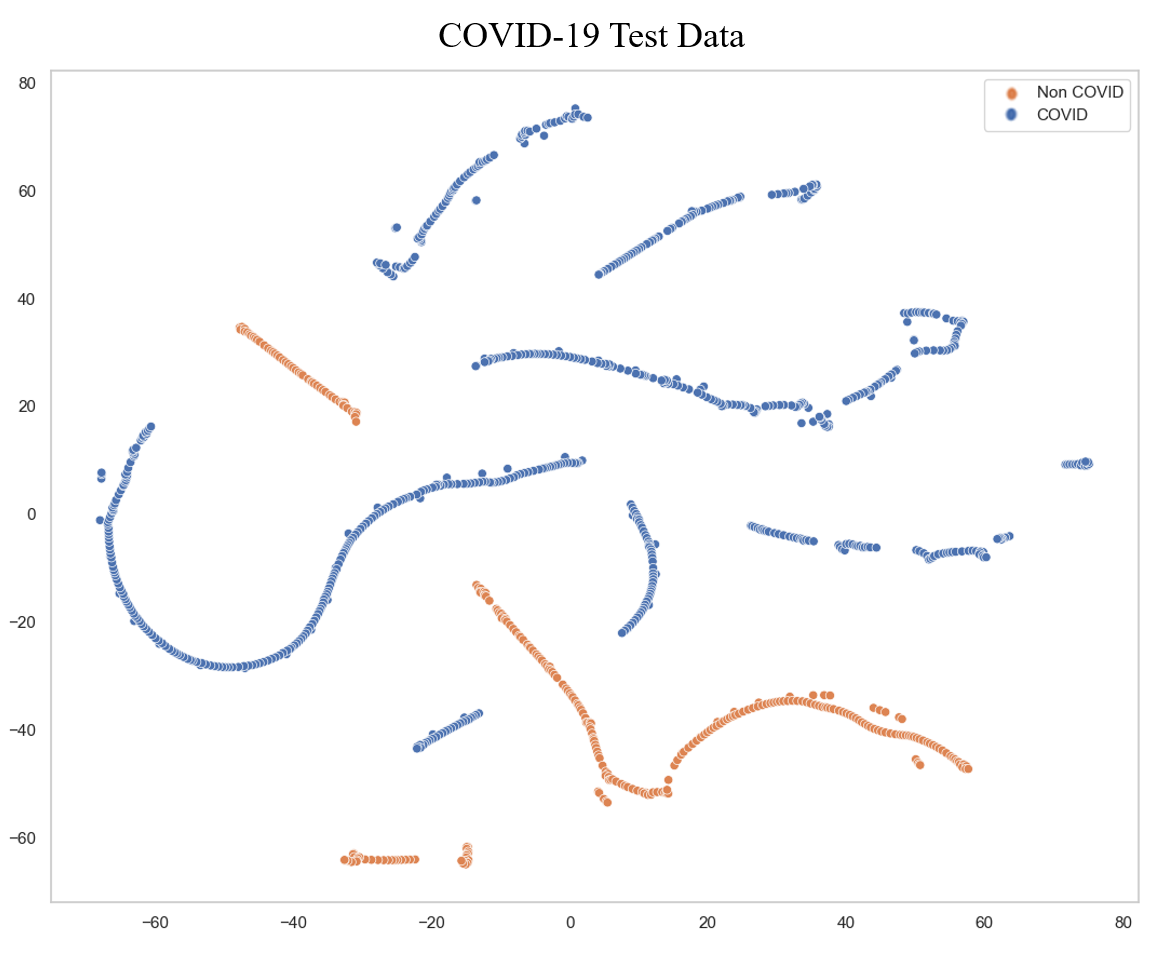}
\end{center}
\caption{{Latent space visualization via t-SNE projection for our method on a test COVID-19 dataset, unseen by SS-StyleGAN.}
}
\label{tsne_testset_covid}
\end{figure}

{\subsection{Generalization to New Data} 
Table \ref{results_test_set_COVID} demonstrates our method's performance and generalization ability on a test COVID-19 dataset, which was not seen before by SS-StyleGAN. The test images were embedded to the latent space using our COVID-19 pretrained SS-StyleGAN where t-SNE, FPS and NN classification were performed. Fig. \ref{tsne_testset_covid} presents a visualization of the t-SNE projected space. The classification results as well as the distinct separation in the latent space show that our self-supervised StyleGAN based pre-training is able to generalize well to new data.}

\begin{table*}[t]
    \caption{{Average classification results on an unseen COVID-19 test dataset with our method when training with 10, 20 and 50 images from all experiments (higher is better).}}
    \label{results_test_set_COVID}
    \centering
    \makebox[\textwidth][c]{
    \begin{tabular}{c|c|c|c|c|c}
    \toprule
    
        \thead{\bf{\normalsize Images}} &  \thead{\bf{\normalsize Accuracy}} &  \thead{\bf{\normalsize Sensitivity}} & \thead{\bf{\normalsize Specificity}} & \thead{\bf{\normalsize Precision}} & \thead{\bf{\normalsize AUC}} \\ \hline
        
        10 &  0.87 $\pm$ 0.07 & 0.78 $\pm$ 0.20 & 0.91  $\pm$ 0.07 & 0.92 $\pm$ 0.06 & 0.84 $\pm$ 0.09\\\hline

        20 & 0.98 $\pm$ 0.02 & 0.95 $\pm$ 0.04 & 0.99  $\pm$ 0.01 & 0.98 $\pm$ 0.02 & 0.97 $\pm$ 0.03\\\hline
        
        50 & 1.00 $\pm$ 0.00 & 1.00 $\pm$ 0.00 & 1.00 $\pm$ 0.00 & 1.00 $\pm$ 0.00 & 1.00 $\pm$ 0.00\\

    \bottomrule
    \end{tabular}}
\end{table*}

\section{Conclusion}
We present SS-StyleGAN, a self-supervised StyleGAN dedicated for image annotation and classification. By leveraging the disentanglement of the StyleGAN latent space, we design a smart representative selection algorithm of samples to be labeled for the classification task which enables the classification with extremely limited training data. 
Our proposed framework incorporates an encoder within the StyleGAN architecture to learn a latent space encouraged to be well suited for inversion while retaining its disentangled properties.  

We demonstrate that our method outperforms state-of-the-art self-supervised learning, supervised image classification, and active learning methods on two medical image datasets. Moreover, we show that using only 10 images for training is sufficient to achieve adequate classification results. 

Future work includes examining iterative sampling of images to be labeled by adopting an active learning approach. Another interesting direction may include manipulation of the latent space to generate class-specific images to further improve the classification results. Furthermore, though we have presented only binary classification tasks, our work can be extended to multi-class classification and SS-StyleGAN can be adopted for other downstream tasks such as detection.  We defer these extensions to future research.

In summary, our proposed framework is generic and can be beneficial to many additional medical tasks and applications by improving classification in scenarios of limited annotated datasets.

{\small
\bibliographystyle{IEEEtran}
\bibliography{bib.bib}

\begin{thebibliography}{10}
\providecommand{\url}[1]{#1}
\csname url@samestyle\endcsname
\providecommand{\newblock}{\relax}
\providecommand{\bibinfo}[2]{#2}
\providecommand{\BIBentrySTDinterwordspacing}{\spaceskip=0pt\relax}
\providecommand{\BIBentryALTinterwordstretchfactor}{4}
\providecommand{\BIBentryALTinterwordspacing}{\spaceskip=\fontdimen2\font plus
\BIBentryALTinterwordstretchfactor\fontdimen3\font minus
  \fontdimen4\font\relax}
\providecommand{\BIBforeignlanguage}[2]{{%
\expandafter\ifx\csname l@#1\endcsname\relax
\typeout{** WARNING: IEEEtran.bst: No hyphenation pattern has been}%
\typeout{** loaded for the language `#1'. Using the pattern for}%
\typeout{** the default language instead.}%
\else
\language=\csname l@#1\endcsname
\fi
#2}}
\providecommand{\BIBdecl}{\relax}
\BIBdecl

\bibitem{frid2018gan}
M.~Frid-Adar, I.~Diamant, E.~Klang, M.~Amitai, J.~Goldberger, and H.~Greenspan,
  ``Gan-based synthetic medical image augmentation for increased cnn
  performance in liver lesion classification,'' \emph{Neurocomputing}, vol.
  321, pp. 321--331, 2018.

\bibitem{zhou2021review}
S.~K. Zhou, H.~Greenspan, C.~Davatzikos, J.~S. Duncan, B.~Van~Ginneken,
  A.~Madabhushi, J.~L. Prince, D.~Rueckert, and R.~M. Summers, ``A review of
  deep learning in medical imaging: Imaging traits, technology trends, case
  studies with progress highlights, and future promises,'' \emph{Proceedings of
  the IEEE}, 2021.

\bibitem{simclr}
T.~Chen, S.~Kornblith, M.~Norouzi, and G.~Hinton, ``A simple framework for
  contrastive learning of visual representations,'' in \emph{International
  conference on machine learning}, 2020, pp. 1597--1607.

\bibitem{moco}
K.~He, H.~Fan, Y.~Wu, S.~Xie, and R.~Girshick, ``Momentum contrast for
  unsupervised visual representation learning,'' in \emph{Proceedings of the
  IEEE/CVF Conference on Computer Vision and Pattern Recognition}, 2020, pp.
  9729--9738.

\bibitem{snail_few_shot}
N.~Mishra, M.~Rohaninejad, X.~Chen, and P.~Abbeel, ``A simple neural attentive
  meta-learner,'' \emph{arXiv preprint arXiv:1707.03141}, 2017.

\bibitem{meta_few_shot}
Q.~Sun, Y.~Liu, T.-S. Chua, and B.~Schiele, ``Meta-transfer learning for
  few-shot learning,'' in \emph{Proceedings of the IEEE/CVF Conference on
  Computer Vision and Pattern Recognition}, 2019, pp. 403--412.

\bibitem{VAAL}
S.~Sinha, S.~Ebrahimi, and T.~Darrell, ``Variational adversarial active
  learning,'' in \emph{Proceedings of the IEEE/CVF International Conference on
  Computer Vision}, 2019, pp. 5972--5981.

\bibitem{coreset}
O.~Sener and S.~Savarese, ``Active learning for convolutional neural networks:
  A core-set approach,'' \emph{arXiv preprint arXiv:1708.00489}, 2017.

\bibitem{AL_dropout}
Y.~Gal and Z.~Ghahramani, ``Dropout as a bayesian approximation: Representing
  model uncertainty in deep learning,'' in \emph{international conference on
  machine learning}.\hskip 1em plus 0.5em minus 0.4em\relax PMLR, 2016, pp.
  1050--1059.

\bibitem{AL_yoo}
D.~Yoo and I.~S. Kweon, ``Learning loss for active learning,'' in
  \emph{Proceedings of the IEEE/CVF Conference on Computer Vision and Pattern
  Recognition}, 2019, pp. 93--102.

\bibitem{AL_combined_adaptive}
X.~Li and Y.~Guo, ``Adaptive active learning for image classification,'' in
  \emph{Proceedings of the IEEE Conference on Computer Vision and Pattern
  Recognition}, 2013, pp. 859--866.

\bibitem{AL_combined_cost}
K.~Wang, D.~Zhang, Y.~Li, R.~Zhang, and L.~Lin, ``Cost-effective active
  learning for deep image classification,'' \emph{IEEE Transactions on Circuits
  and Systems for Video Technology}, vol.~27, no.~12, pp. 2591--2600, 2016.

\bibitem{11_1_gidaris2018unsupervised}
S.~Gidaris, P.~Singh, and N.~Komodakis, ``Unsupervised representation learning
  by predicting image rotations,'' in \emph{ICLR 2018}, 2018.

\bibitem{14_1_goyal2019scaling}
P.~Goyal, D.~Mahajan, A.~Gupta, and I.~Misra, ``Scaling and benchmarking
  self-supervised visual representation learning,'' in \emph{Proceedings of the
  IEEE/CVF International Conference on Computer Vision}, 2019, pp. 6391--6400.

\bibitem{mocov2}
X.~Chen, H.~Fan, R.~Girshick, and K.~He, ``Improved baselines with momentum
  contrastive learning,'' \emph{arXiv preprint arXiv:2003.04297}, 2020.

\bibitem{49_1_wei2018learning}
D.~Wei, J.~J. Lim, A.~Zisserman, and W.~T. Freeman, ``Learning and using the
  arrow of time,'' in \emph{CVPR}, 2018, pp. 8052--8060.

\bibitem{4_1_chen2019self}
T.~Chen, X.~Zhai, M.~Ritter, M.~Lucic, and N.~Houlsby, ``Self-supervised gans
  via auxiliary rotation loss,'' in \emph{Proceedings of the IEEE/CVF
  Conference on Computer Vision and Pattern Recognition}, 2019, pp.
  12\,154--12\,163.

\bibitem{lt-gan}
P.~Patel, N.~Kumari, M.~Singh, and B.~Krishnamurthy, ``Lt-gan: Self-supervised
  gan with latent transformation detection,'' in \emph{Proceedings of the
  IEEE/CVF Winter Conference on Applications of Computer Vision}, 2021, pp.
  3189--3198.

\bibitem{27_1_luvcic2019high}
M.~Lu{\v{c}}i{\'c}, M.~Tschannen, M.~Ritter, X.~Zhai, O.~Bachem, and S.~Gelly,
  ``High-fidelity image generation with fewer labels,'' in \emph{International
  Conference on Machine Learning}.\hskip 1em plus 0.5em minus 0.4em\relax PMLR,
  2019, pp. 4183--4192.

\bibitem{gmm-gan-yuri}
Y.~Feigin, H.~Spitzer, and R.~Giryes, ``Gmm-based generative adversarial
  encoder learning,'' \emph{arXiv preprint arXiv:2012.04525}, 2020.

\bibitem{51_1_zhang2019aet}
L.~Zhang, G.-J. Qi, L.~Wang, and J.~Luo, ``{AET} vs. {AED}: Unsupervised
  representation learning by auto-encoding transformations rather than data,''
  in \emph{CVPR}, 2019, pp. 2547--2555.

\bibitem{karras2019style}
T.~Karras, S.~Laine, and T.~Aila, ``A style-based generator architecture for
  generative adversarial networks,'' in \emph{Proceedings of the IEEE/CVF
  Conference on Computer Vision and Pattern Recognition}, 2019, pp. 4401--4410.

\bibitem{stylegan2}
T.~Karras, S.~Laine, M.~Aittala, J.~Hellsten, J.~Lehtinen, and T.~Aila,
  ``Analyzing and improving the image quality of stylegan,'' in
  \emph{Proceedings of the IEEE/CVF Conference on Computer Vision and Pattern
  Recognition}, 2020, pp. 8110--8119.

\bibitem{abdal2019image2stylegan}
R.~Abdal, Y.~Qin, and P.~Wonka, ``Image2stylegan: How to embed images into the
  stylegan latent space?'' in \emph{Proceedings of the IEEE international
  conference on computer vision}, 2019, pp. 4432--4441.

\bibitem{hochberg2021style}
D.~C. Hochberg, R.~Giryes, and H.~Greenspan, ``Style encoding for
  class-specific image generation,'' in \emph{Medical Imaging 2021: Image
  Processing}, vol. 11596.\hskip 1em plus 0.5em minus 0.4em\relax International
  Society for Optics and Photonics, 2021, p. 1159631.

\bibitem{pidhorskyi2020adversarial}
S.~Pidhorskyi, D.~A. Adjeroh, and G.~Doretto, ``Adversarial latent
  autoencoders,'' in \emph{Proceedings of the IEEE/CVF Conference on Computer
  Vision and Pattern Recognition}, 2020, pp. 14\,104--14\,113.

\bibitem{richardson2020encoding}
E.~Richardson, Y.~Alaluf, O.~Patashnik, Y.~Nitzan, Y.~Azar, S.~Shapiro, and
  D.~Cohen-Or, ``Encoding in style: a stylegan encoder for image-to-image
  translation,'' \emph{arXiv preprint arXiv:2008.00951}, 2020.

\bibitem{FPS}
P.~Kamousi, S.~Lazard, A.~Maheshwari, and S.~Wuhrer, ``Analysis of farthest
  point sampling for approximating geodesics in a graph,'' \emph{Computational
  Geometry}, vol.~57, pp. 1--7, 2016.

\bibitem{chen2020big}
T.~Chen, S.~Kornblith, K.~Swersky, M.~Norouzi, and G.~E. Hinton, ``Big
  self-supervised models are strong semi-supervised learners,'' \emph{Advances
  in Neural Information Processing Systems}, vol.~33, pp. 22\,243--22\,255,
  2020.

\bibitem{self-supervised-medical}
S.~Azizi, B.~Mustafa, F.~Ryan, Z.~Beaver, J.~Freyberg, J.~Deaton, A.~Loh,
  A.~Karthikesalingam, S.~Kornblith, T.~Chen, V.~Natarajan, and M.~Norouzi,
  ``Big self-supervised models advance medical image classification,'' in
  \emph{Proceedings of the IEEE/CVF International Conference on Computer Vision
  (ICCV)}, October 2021, pp. 3478--3488.

\bibitem{sowrirajan2020moco}
H.~Sowrirajan, J.~Yang, A.~Y. Ng, and P.~Rajpurkar, ``Moco-cxr: Moco
  pretraining improves representation and transferability of chest x-ray
  models,'' in \emph{Medical Imaging with Deep Learning (MIDL) Conference},
  2021.

\bibitem{zhou2020comparing}
H.-Y. Zhou, S.~Yu, C.~Bian, Y.~Hu, K.~Ma, and Y.~Zheng, ``Comparing to learn:
  Surpassing imagenet pretraining on radiographs by comparing image
  representations,'' in \emph{MICCAI}.\hskip 1em plus 0.5em minus 0.4em\relax
  Springer, 2020, pp. 398--407.

\bibitem{spitzer2018improving}
H.~Spitzer, K.~Kiwitz, K.~Amunts, S.~Harmeling, and T.~Dickscheid, ``Improving
  cytoarchitectonic segmentation of human brain areas with self-supervised
  siamese networks,'' in \emph{International Conference on Medical Image
  Computing and Computer-Assisted Intervention}.\hskip 1em plus 0.5em minus
  0.4em\relax Springer, 2018, pp. 663--671.

\bibitem{bai2019self}
W.~Bai, C.~Chen, G.~Tarroni, J.~Duan, F.~Guitton, S.~E. Petersen, Y.~Guo, P.~M.
  Matthews, and D.~Rueckert, ``Self-supervised learning for cardiac mr image
  segmentation by anatomical position prediction,'' in \emph{International
  Conference on Medical Image Computing and Computer-Assisted
  Intervention}.\hskip 1em plus 0.5em minus 0.4em\relax Springer, 2019, pp.
  541--549.

\bibitem{goodfellow2014generative}
I.~Goodfellow, J.~Pouget-Abadie, M.~Mirza, B.~Xu, D.~Warde-Farley, S.~Ozair,
  A.~Courville, and Y.~Bengio, ``Generative adversarial nets,'' \emph{Advances
  in neural information processing systems}, vol.~27, pp. 2672--2680, 2014.

\bibitem{karras2018progressive}
T.~Karras, T.~Aila, S.~Laine, and J.~Lehtinen, ``Progressive growing of gans
  for improved quality, stability, and variation,'' in \emph{International
  Conference on Learning Representations}, 2018.

\bibitem{yang2021semantic}
C.~Yang, Y.~Shen, and B.~Zhou, ``Semantic hierarchy emerges in deep generative
  representations for scene synthesis,'' \emph{International Journal of
  Computer Vision}, vol. 129, no.~5, pp. 1451--1466, 2021.

\bibitem{iGAN}
J.-Y. Zhu, P.~Kr{\"a}henb{\"u}hl, E.~Shechtman, and A.~A. Efros, ``Generative
  visual manipulation on the natural image manifold,'' in \emph{European
  conference on computer vision}.\hskip 1em plus 0.5em minus 0.4em\relax
  Springer, 2016, pp. 597--613.

\bibitem{BiGAN}
J.~Donahue, P.~Kr{\"a}henb{\"u}hl, and T.~Darrell, ``Adversarial feature
  learning,'' \emph{arXiv preprint arXiv:1605.09782}, 2016.

\bibitem{dcgan}
A.~Radford, L.~Metz, and S.~Chintala, ``Unsupervised representation learning
  with deep convolutional generative adversarial networks,'' \emph{arXiv
  preprint arXiv:1511.06434}, 2015.

\bibitem{e4e_encoder}
O.~Tov, Y.~Alaluf, Y.~Nitzan, O.~Patashnik, and D.~Cohen-Or, ``Designing an
  encoder for stylegan image manipulation,'' \emph{ACM Transactions on Graphics
  (TOG)}, vol.~40, no.~4, pp. 1--14, 2021.

\bibitem{face_stylegan_enc}
Y.~Nitzan, A.~Bermano, Y.~Li, and D.~Cohen-Or, ``Face identity disentanglement
  via latent space mapping,'' \emph{ACM Transactions on Graphics (TOG)},
  vol.~39, pp. 1 -- 14, 2020.

\bibitem{guan2020collaborative}
S.~Guan, Y.~Tai, B.~Ni, F.~Zhu, F.~Huang, and X.~Yang, ``Collaborative learning
  for faster stylegan embedding,'' \emph{arXiv preprint arXiv:2007.01758},
  2020.

\bibitem{AL_synthesis1}
D.~Mahapatra, B.~Bozorgtabar, J.-P. Thiran, and M.~Reyes, ``Efficient active
  learning for image classification and segmentation using a sample selection
  and conditional generative adversarial network,'' in \emph{International
  Conference on Medical Image Computing and Computer-Assisted
  Intervention}.\hskip 1em plus 0.5em minus 0.4em\relax Springer, 2018, pp.
  580--588.

\bibitem{AL_sampling}
S.~Burr \emph{et~al.}, ``Active learning,'' \emph{Synthesis Lectures on
  Artificial Intelligence and Machine Learning}, vol.~6, no.~1, pp. 1--114,
  2012.

\bibitem{AL_diversity}
Y.~Yang, Z.~Ma, F.~Nie, X.~Chang, and A.~G. Hauptmann, ``Multi-class active
  learning by uncertainty sampling with diversity maximization,''
  \emph{International Journal of Computer Vision}, vol. 113, no.~2, pp.
  113--127, 2015.

\bibitem{AL_clustering}
H.~T. Nguyen and A.~Smeulders, ``Active learning using pre-clustering,'' in
  \emph{Proceedings of the twenty-first international conference on Machine
  learning}, 2004, p.~79.

\bibitem{AL_high}
D.~L. Donoho \emph{et~al.}, ``High-dimensional data analysis: The curses and
  blessings of dimensionality,'' \emph{AMS math challenges lecture}, vol.~1,
  no. 2000, p.~32, 2000.

\bibitem{tsne}
L.~Van~der Maaten and G.~Hinton, ``Visualizing data using t-sne.''
  \emph{Journal of machine learning research}, vol.~9, no.~11, 2008.

\bibitem{nearest_neighbor}
T.~Cover and P.~Hart, ``Nearest neighbor pattern classification,'' \emph{IEEE
  transactions on information theory}, vol.~13, no.~1, pp. 21--27, 1967.

\bibitem{FPS_image}
Y.~Eldar, M.~Lindenbaum, M.~Porat, and Y.~Y. Zeevi, ``The farthest point
  strategy for progressive image sampling,'' \emph{IEEE Transactions on Image
  Processing}, vol.~6, no.~9, pp. 1305--1315, 1997.

\bibitem{PCA}
I.~T. Jolliffe and J.~Cadima, ``Principal component analysis: a review and
  recent developments,'' \emph{Philosophical Transactions of the Royal Society
  A: Mathematical, Physical and Engineering Sciences}, vol. 374, no. 2065, p.
  20150202, 2016.

\bibitem{tsne_face}
J.~Yi, X.~Mao, Y.~Xue, and A.~Compare, ``Facial expression recognition based on
  t-sne and adaboostm2,'' in \emph{2013 IEEE International Conference on Green
  Computing and Communications and IEEE Internet of Things and IEEE Cyber,
  Physical and Social Computing}.\hskip 1em plus 0.5em minus 0.4em\relax IEEE,
  2013, pp. 1744--1749.

\bibitem{tsne_tumor}
W.~M. Abdelmoula, B.~Balluff, S.~Englert, J.~Dijkstra, M.~J. Reinders,
  A.~Walch, L.~A. McDonnell, and B.~P. Lelieveldt, ``Data-driven identification
  of prognostic tumor subpopulations using spatially mapped t-sne of mass
  spectrometry imaging data,'' \emph{Proceedings of the National Academy of
  Sciences}, vol. 113, no.~43, pp. 12\,244--12\,249, 2016.

\bibitem{covid_data_cheng}
J.~Ma, Y.~Wang, X.~An, C.~Ge, Z.~Yu, J.~Chen, Q.~Zhu, G.~Dong, J.~He, Z.~He,
  T.~Cao, Y.~Zhu, Z.~Nie, and X.~Yang, ``Toward data-efficient learning: A
  benchmark for covid-19 ct lung and infection segmentation,'' \emph{Medical
  physics}, vol.~48, no.~3, p. 1197—1210, March 2021.

\bibitem{covid_joseph}
\BIBentryALTinterwordspacing
J.~P. Cohen, P.~Morrison, and L.~Dao, ``Covid-19 image data collection,''
  \emph{arXiv 2003.11597}, 2020. [Online]. Available:
  \url{https://github.com/ieee8023/covid-chestxray-dataset}
\BIBentrySTDinterwordspacing

\bibitem{LiTS}
C.~P, E.~F, L.~J, and K.~G., ``{LiTS - Liver Tumor Segmentation Challenge},''
  \url{http://www.lits-challenge.com/}, 2017.

\bibitem{efficientnet}
M.~Tan and Q.~Le, ``Efficientnet: Rethinking model scaling for convolutional
  neural networks,'' in \emph{ICML}.\hskip 1em plus 0.5em minus 0.4em\relax
  PMLR, 2019, pp. 6105--6114.

\bibitem{deng2009imagenet}
J.~Deng, W.~Dong, R.~Socher, L.-J. Li, K.~Li, and L.~Fei-Fei, ``Imagenet: A
  large-scale hierarchical image database,'' in \emph{2009 IEEE conference on
  computer vision and pattern recognition}.\hskip 1em plus 0.5em minus
  0.4em\relax Ieee, 2009, pp. 248--255.

\bibitem{moco_pretraining}
H.~Sowrirajan, J.~Yang, A.~Y. Ng, and P.~Rajpurkar, ``Moco pretraining improves
  representation and transferability of chest x-ray models,'' in \emph{Medical
  Imaging with Deep Learning}.\hskip 1em plus 0.5em minus 0.4em\relax PMLR,
  2021, pp. 728--744.

\bibitem{covid_dataset_unseen}
K.~Zhang, X.~Liu, J.~Shen, Z.~Li, Y.~Sang, X.~Wu, Y.~Zha, W.~Liang, C.~Wang,
  K.~Wang \emph{et~al.}, ``Clinically applicable ai system for accurate
  diagnosis, quantitative measurements, and prognosis of covid-19 pneumonia
  using computed tomography,'' \emph{Cell}, vol. 181, no.~6, pp. 1423--1433,
  2020.

\bibitem{FID}
M.~Heusel, H.~Ramsauer, T.~Unterthiner, B.~Nessler, and S.~Hochreiter, ``Gans
  trained by a two time-scale update rule converge to a local nash
  equilibrium,'' \emph{Advances in neural information processing systems},
  vol.~30, 2017.

\bibitem{inception}
K.~Simonyan and A.~Zisserman, ``Very deep convolutional networks for
  large-scale image recognition,'' \emph{arXiv preprint arXiv:1409.1556}, 2014.

\end{thebibliography}
}

\end{document}